\documentclass[aps,prl,twocolumn,superscriptaddress,showpacs]{revtex4-1}
\usepackage{graphicx}% Include Fig. files
\usepackage{epstopdf}
\usepackage{dcolumn}% Align table columns on decimal point
\usepackage{bm}% bold math
\usepackage{amssymb}
\usepackage{txfonts}
\usepackage{bm}% bold math
\usepackage{color}
\usepackage{upgreek}

\begin{document}

%\preprint{}

%
\title{CRYSTAL STRUCTURE, INCOMMENSURATE MAGNETIC ORDER AND FERROELECTRICITY IN Mn$_{1-x}$Cu$_{x}$WO${_4}$ (x=0 -- 0.19)}
\author
{C. M. N. Kumar}
\email{n.kumar@fz-juelich.de}
\email{naveenkumarcm@gmail.com}
\affiliation{Forschungszentrum J\"{u}lich GmbH, J\"{u}lich Centre for Neutron Science (JCNS), Oak Ridge National Laboratory, Oak Ridge, Tennessee 37831, United States}
\affiliation{Chemical and Engineering Materials Division, Oak Ridge National Laboratory, Oak Ridge, Tennessee 37831, United States}
\author{Y. Xiao}
\affiliation{Forschungszentrum J\"{u}lich GmbH, J\"{u}lich Centre for Neutron Science (JCNS) and Peter Gr\"{u}nberg Institut PGI, Forschungszentrum J\"{u}lich, D-52425 J\"{u}lich, Germany}
\author{P. Lunkenheimer}
\affiliation{Experimental Physics V, Center for Electronic Correlations and Magnetism, University of Augsburg, 86135 Augsburg, Germany}
\author{A. Loidl}
\affiliation{Experimental Physics V, Center for Electronic Correlations and Magnetism, University of Augsburg, 86135 Augsburg, Germany}
\author{M. Ohl}
\affiliation{Forschungszentrum J\"{u}lich GmbH, J\"{u}lich Centre for Neutron Science (JCNS), Oak Ridge National Laboratory, Oak Ridge, 37831, United States}
\affiliation{Biology and Soft Matter Division, Oak Ridge National Laboratory, Oak Ridge, Tennessee 37831, United States}
\affiliation{Department of Material Science and Engineering, University of Tennessee, Knoxville, Tennessee 37996, United States}
\date{\today}
\begin{abstract}
We have carried out a systematic study on the effect of Cu doping on nuclear, magnetic, and dielectric properties in Mn$_{1-x}$Cu$_{x}$WO$_4$ for ${0}\leq{x}\leq{0.19}$ by a synergic use of different techniques, viz, heat capacity, magnetization, dielectric, and neutron powder diffraction measurements. Via heat capacity and magnetization measurements we show that with increasing Cu concentration magnetic frustration decreases, which leads to the stabilization of commensurate magnetic ordering. This was further verified by temperature-dependent unit cell volume changes derived from neutron diffraction measurements which was modeled by the Gr\"{u}neisen approximation. Dielectric measurements show a low temperature phase transition below about 9-10~K. Further more, magnetic refinements reveal no changes below this transition indicating a possible spin-flop transition which is unique to the Cu doped system. From these combined studies we have constructed a magnetoelectric phase diagram of this compound.
\end{abstract}
\maketitle
%
%%%%%%%%%%%%%%%%%%%%%%%%%%%%%%%%%%%%%%%%%%%%%%%%%%%%%%%%%%%%%%%%%%%%%%%%%%%%%%%%%%%%%%%%%%%%
%
\section{I. Introduction}
A typical feature of multiferroic materials undergoing a transition to an elliptic spiral ferroelectric phase, is the existence of spectacular magnetoelectric effects, such as the polarization flops observed in TbMnO$_3$~\cite{TKimuranature_426Nature2003} and orthorhombic DyMnO$_3$~\cite{JStrempfer_75PRB2007absence} or the sign reversal of electric polarization which is revealed under magnetic field in TbMn$_2$O$_5$~\cite{NHur_392Nature2004}. The orientation of the applied magnetic field with respect to the magnetic spins influences the stability range of the spiral phase and the electric polarization-flop process. This property was recently illustrated by remarkable magnetic field induced effects observed in ferroelectric phase of manganese tungstate MnWO$_4$ in which applied field induces a polarization flop transition~\cite{GLoutenschaelger_48PRB1993Magnetic, KTaniguchi_97PRL2006ferroelectric, HSagayama_77PRB2008Correlation, KTaniguchi_101PRL2008Control}.

In most of the recently discovered multiferroics, the ferroelectric polarization can be explained by the inverse Dzyaloshinski-Moriya effect~\cite{HKatsura_98PRL2007dynamical, MMostovoy_96PRL2006ferroelectricity, TKimura_71PRB2005magnetoelectric}, where the induced electric polarization of a single pair of spins $\vec{S}_i,~\vec{S}_j$ separated by a distance vector $\vec{r}_{i,j}$ is given by~\cite{HKatsura_98PRL2007dynamical}

\begin{equation}
\label{Polarization_spin_relation}
{\vec{P}_{FE}} \propto {\vec{r}_{ij}} \times \left( {\vec{S_i} \times \vec{S_j}} \right).
\end{equation}

The required helical magnetic structure may arise from strong frustration. Since in addition the interaction, equation~\ref{Polarization_spin_relation}, is only a second order effect, the ferroelectric polarization is rather small in these materials. In the RMnO$_3$~\cite{TKimuranature_426Nature2003, TGoto_92PRL2004ferroelectricity} (R=rare earth) series and in MnWO$_4$~\cite{KTaniguchi_97PRL2006ferroelectric, AHArkenbout_74PRB2006ferroelectricity, OHeyer_18JPCM2006Anew} the electric polarization is about two to three orders of magnitude smaller than in a classical ferroelectric perovskite such as BaTiO$_3$. As a consequence the observation of electric--field--induced effects in the magnetically ordered state is more difficult. Nevertheless, it was shown that it is possible in these chiral multiferroics to switch the magnetic order by the application of an electric field at constant temperature~\cite{TLottermoser_430Nature2004magnetic,TFinger_81PRB2010ElectricField, APoole_145JPCS2009}.

The crystal structure of MnWO$_4$ is monoclinic with space group $P2/c$, made up of MnO$_6$ octahedra with high-spin Mn$^{2+}$ ($d_5$) ions and WO$_6$ octahedra with diamagnetic W$^{6+}$ ($d_0$) ions~\cite{HWeitzel_7SSC1969TwoAntiferromagnetic}. Recently it was found that MnWO$_4$ exhibits multiferroicity in which magnetism causes ferroelectricity, implying a strong coupling between the two~\cite{OHeyer_18JPCM2006Anew, KTaniguchi_97PRL2006ferroelectric, AHArkenbout_74PRB2006ferroelectricity}. MnWO$_4$ is one of the prototypical multiferroic materials exhibiting spin-current ferroelectricity~\cite{KTaniguchi_97PRL2006ferroelectric}. It possesses a complex phase diagram with $3$ antiferromagnetic phases below $14$~K namely AF1, AF2 and AF3 at zero magnetic field. AF2 is a ferroelectric (FE) phase, in which the net polarization is along the $b$ axis which can be flipped to the $a$ axis with the application of an external magnetic field. This is the first example of the ferroelectric polarization flop induced by magnetic fields in transition-metal oxide systems without rare-earth $4f$ moments. Taniguchi~$et~al.$ showed that the stability of the magnetoelectric domain walls in a canted magnetic field plays a key role in the directional control of the electric polarization flop phenomenon~\cite{KTaniguchi_101PRL2008Control}. From polarized neutron scattering measurements Sagayama~$et~al.$ showed that an inverse effect of Dzyaloshinskii-Moriya interaction is the origin of the spontaneous electric polarization in the spiral phase of MnWO$_4$~\cite{HSagayama_77PRB2008Correlation}. From superspace symmetry formalism it was shown that in the AF3 phase, the modulations of two Mn atoms within the unit cell can have a cycloidal component with equal and opposite chiralities canceling their effects and hence no electric polarization is induced. Whereas in the AF2 phase, an additional second magnetic mode with the spin modulations breaks the symmetry relation between the two manganese atoms with chiralities of the same sign which add up to induce macroscopic electric polarization~\cite{Iurcelay_87PRB2013incommensurate}.

Recently it was found that the ferroelectric phase is completely suppressed in MnWO$_4$ by doping $10$\% iron on Mn site, which can be again restored with the application of a magnetic field. The absence of ferroelectricity (at zero field) in Mn$_{0.9}$Fe$_{0.1}$WO$_{4}$ is explained by the increase of uniaxial spin anisotropy $K$~\cite{RPChaudhury_77PRB2008suppression}. Evidence for the increase of $K$ with Fe substitution was also derived from neutron scattering experiments~\cite{EGarciaMatres_32EPJ2003magnetic}. It was observed that in Mn$_{1-x}$ M$_{x}$WO$_4$ (M=Mg, Zn and $x\leq0.3$), the substitution of the nonmagnetic Mg$^{2+}$ ions and Zn$^{2+}$ for the magnetic Mn$^{2+}$ ions result in very similar effects on the magnetic and dielectric properties of MnWO$_4$~\cite{LMeddar_21CM2009effect}. These substitutions destabilized the non-polar magnetic structure AF1 of MnWO$_4$ but the AF3-to-AF2 magnetoelectric phase transition was not affected. This indicated that the nonmagnetic dopant destroys neither the three-dimensional nature of magnetic interactions, nor the spin frustration within each $\parallel c$--chain and between $\parallel c$--chains along the $a$--direction. In this article we discuss the influence of doping of Cu ions on the nuclear and magnetic structure of MnWO$_4$.

\section{II. Experimental}
Polycrystalline powders of Mn$_{1-x}$Cu$_{x}$WO$_{4}$ ($x$=0.0-0.19) were prepared by conventional solid state route. Stoichiometric amount of precursors, W$_2$O$_3(99.9 \%)$, MnO$_2(99.9\%)$ and CuO$(99.99\%)$ were ground well with a mortar and pestle, pressed into pellets and sintered in a furnace at $900~^{\circ}C$ for 12 hours in the presence of atmospheric air. This process is repeated to achieve homogeneous powder samples. All compositions were confirmed to be phase pure from x-ray powder diffraction. Specific heat measurements were carried out on small pellets, using a physical property measurement system (Quantum Design) in the temperature range $3$--$300$~K. Magnetic measurements were carried out in a commercial physical property measurement system using vibrating sample magnetometer option. To investigate nuclear and magnetic structure, time--of--flight (TOF) neutron powder diffraction (NPD) was performed on 8~g of powder samples that were loaded in 8~mm diameter vanadium cans. Neutron data were collected at the Spallation Neutron Source (SNS) at Oak Ridge National Laboratory on the high resolution neutron powder diffractometer POWGEN~\cite{AHuq_1ZK2011AThirdGeneration}. Data were collected for the compositions of Cu, x=0.0, 0.05, 0.1 and 0.19 in the temperature range $1.5-300$~K. For each temperature the data was collected using two different center wavelengths, 1.599~{\AA} and 3.731~{\AA} or 4.797~{\AA}. The crystal and magnetic structure refinements were carried out from the NPD data using the Rietveld refinement program $FullProf$~\cite{IRCarvajal_192PhysicaB1993RecentAdvances}. For the dielectric measurements, opposite sides of pressed pellets were covered by silver paint, thus forming a parallel-plate capacitor. The measurements were done using an LCR meter (Agilent~4980). For cooling down to 5~K, a He bath cryostat was used.

\section{III. Results and discussion}

\begin{figure}[!b]
\centering
\includegraphics[scale=0.30]{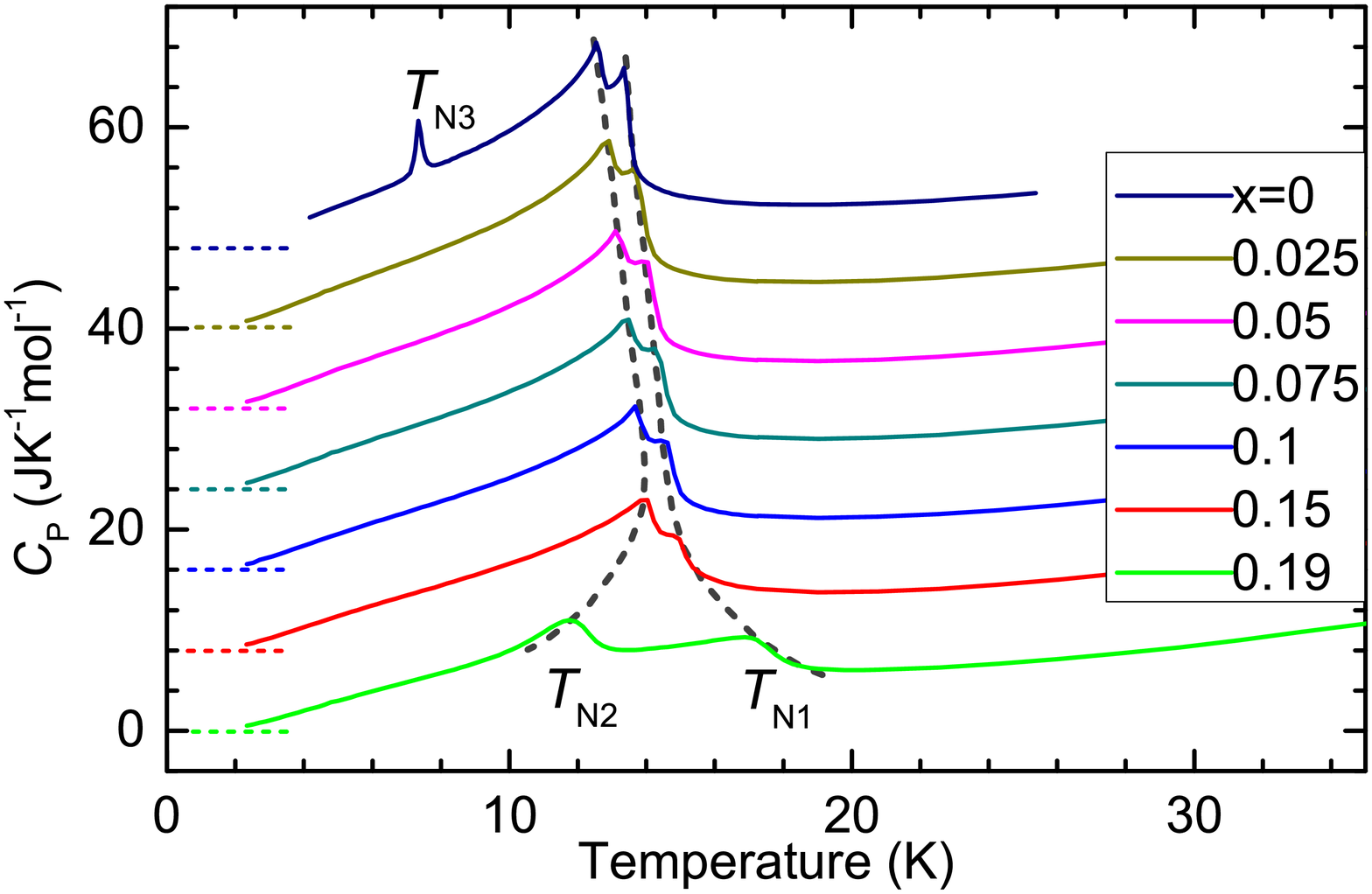}
\caption{(Colour online) The temperature dependence of specific heat $C_\mathrm P$ of Mn$_{1-x}$Cu$_{x}$WO$_{4}$. Different curves are vertically offset by 8 units along $C_{\mathrm {P}}$-axis, zero for each curve is defined by the horizontal dashed lines. Vertical dashed lines are guide to eyes indicating transitions $T_{\mathrm {N1}}$ and $T_{\mathrm {N2}}$.}
\label{Specific_heat}
\end{figure}

The thermodynamic signature of the transition between different phases is usually detected by distinct anomalies in the temperature dependence of the heat capacity, $(C_\mathrm P)$. Multiferroic materials with several subsequent transitions may show pronounced anomalies of $C_\mathrm P$. In figure~\ref{Specific_heat} we present the variation of $C_\mathrm P$ with the temperature for Mn$_{1-x}$Cu$_{x}$WO$_{4}$. For reference, the specific heat of a MnWO$_4$ single crystal is also included. All the compositions show two anomalies at $T_{\mathrm {N1}}$ and $T_{\mathrm {N2}}$. A third low temperature phase transition, $T_{\mathrm {N3}}$ seen in the case of MnW$O_4$ is already missing in the lowest Cu doped compound. This is associated with the phase transition from the helical AF2 phase to the commensurate AF1 phase~\cite{GLoutenschaelger_48PRB1993Magnetic}. This result implies that with Cu doping a quick suppression of the AF1 phase occurs, as a result ferroelectric AF2 phase is extended to the lowest temperature. Similar results of quick suppression of the AF1 phase were reported in Mn$_{1-x}$Co$_{x}$WO$_{4}$ and Mn$_{1-x}$Zn$_{x}$WO$_{4}$~\cite{YSSong_22PRBStabilization,RPChaudhury_83PRB2011robust}.

\begin{figure}[!t]
\centering
\includegraphics[scale=0.7]{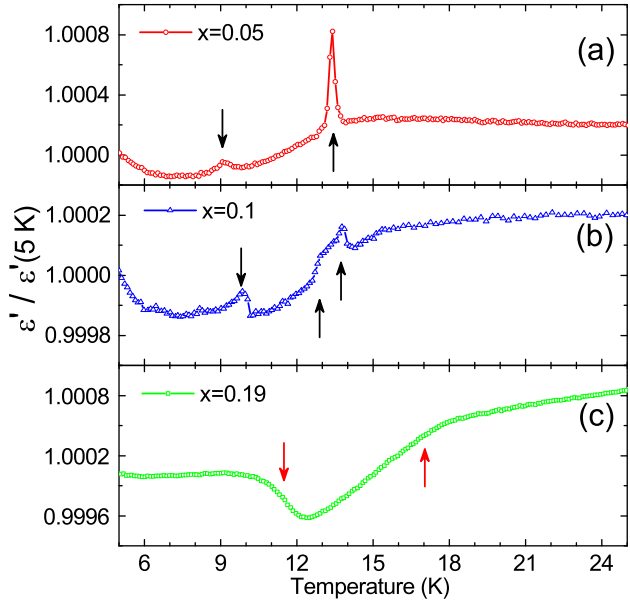}
\caption{(Colour online) Temperature dependence of dielectric constant at 105~kHz, for Mn$_{1-x}$Cu$_{x}$WO$_4$ with: (a) x=0.05, (b) x=0.1 and (c) x=0.19. The arrows in (a) and (b) indicate the dielectric anomalies and arrows in (c) indicate the anomalies seen in specific heat measurement of the sample with x=0.19.}
\label{DielectricPlots}
\end{figure}

\indent
Dielectric measurements were performed on compositions x = 0.05, 0.1, and 0.19. The temperature dependence of the dielectric constant ($\varepsilon^{\prime}$), normalized to the dielectric constant value at 5 K, is presented in figure~\ref{DielectricPlots}. To exclude contributions from electrode polarization or grain boundaries, which can lead to so-called Maxwell-Wagner relaxations, here we show the results at a relatively high frequency of 105~kHz~\cite{PLunkenheimer_66PRB2002Origin,PLunkenheimer_180EPJST2010Colossal}. For x = 0.05 and 0.1, several anomalies in $\varepsilon^{\prime}$(T) are found as indicated by the arrows in figures~\ref{DielectricPlots}(a) and (b). Those around 12~K agree with the findings from specific heat (Fig.~\ref{Specific_heat}). With increased Cu concentration, these dielectric-constant anomalies become weaker. In addition to the two transitions observed from specific-heat measurements, a third transition is revealed by the dielectric measurements at $T_{\mathrm {x}}=9$ and 10~K for the x=0.05 and 0.1 compounds, respectively. From neutron diffraction measurements we later show that $T_{\mathrm {x}}$ is not associated with the phase transition from AF2 to AF1 phase as in parent MnWO$_4$. As indicated by the arrows in figure~\ref{DielectricPlots}(c), the two broad shoulders in $\varepsilon^{\prime}$(T) found for x=0.19 seem to roughly agree with the specific-heat results but a clear determination of transition temperatures from the dielectric experiments is not possible for this sample.

\begin{figure}[!t]
\centering
\includegraphics[scale=0.35]{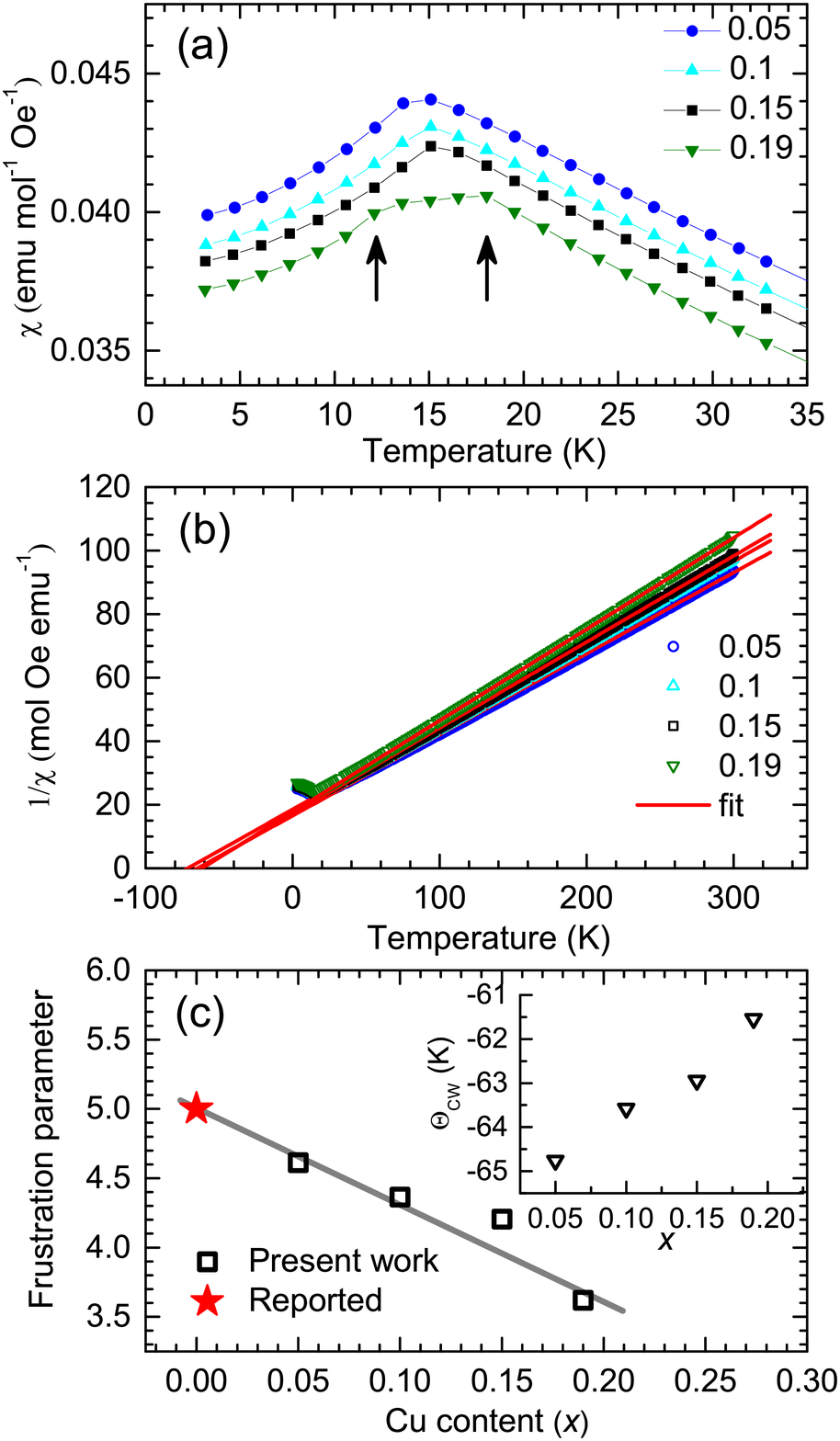}
\caption{(Colour online) (a) Susceptibility calculated from the magnetization data measured after a field cooling cycle with applied magnetic field of 1 kOe. (b) Curie-Weiss fit (red lines) to the inverse susceptibilities, a deviation from linear nature is seen below magnetic ordering temperature.(c) The frustration parameter as a function of composition. Inset shows the Curie-Weiss temperature as a function of composition obtained from the Curie-Weiss fits. Straight lines are guide to eyes.}
\label{Merged_magnetization_properties}
\end{figure}

\begin{figure}[!t]
\centering
\includegraphics[scale=0.45]{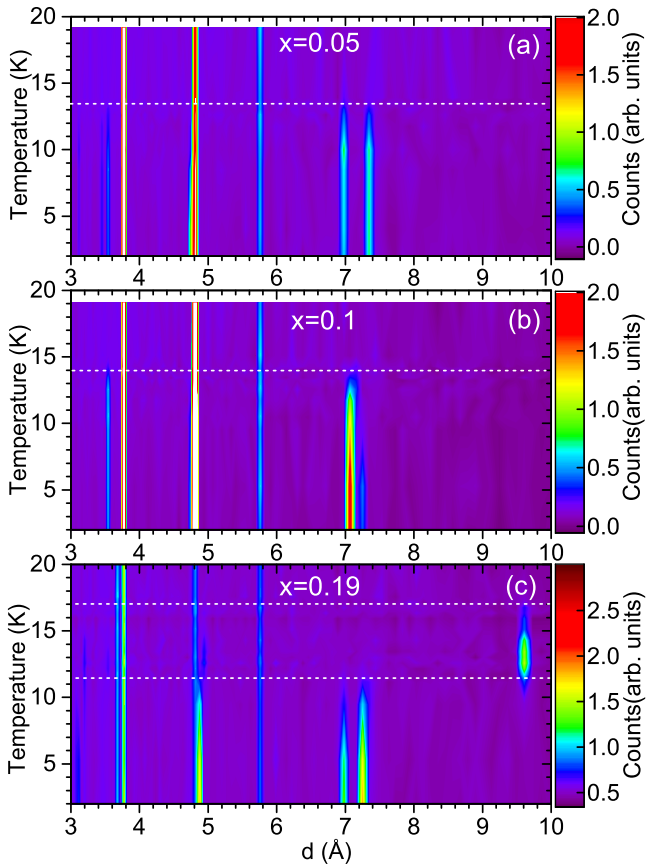}
\caption{(Colour online) (a) Temperature evolution of diffraction patterns of Mn$_{0.95}$Cu$_{0.05}$WO$_4$. Magnetic phase transition is discernible with the appearance of additional incommensurate Bragg peaks around 13.5~K (horizontal dashed line). (b) Temperature evolution of part of the diffraction patterns of Mn$_{0.9}$Cu$_{0.1}$WO$_4$. Magnetic phase transition is discernible with the appearance of additional incommensurate Bragg peaks around 14~K. (c) Temperature dependence of diffraction patterns of Mn$_{0.8}$Cu$_{0.19}$WO$_4$. Two magnetic phase transitions are discernible at $\sim$17~K and $\sim$11.5~K.}
\label{2D_Intensity_Maps}
\end{figure}

\begin{table}[!h]
\caption{Structural data for MnWO$_{4}$, Mn$_{0.9}$Cu$_{0.1}$WO$_{4}$ and Mn$_{0.81}$Cu$_{0.19}$WO$_{4}$ obtained from the NPD data collected at 300~K.}
\label{Refinement_results_three_compositions}
\begin{tabular*}{\columnwidth}{@{\extracolsep{\fill}}ccccc}
  \hline \hline
    &   &  \multicolumn{3}{c}{Cu content (x)}\\ \cline{3-5} %% This will add a horizontal line along the cells 3 to 5
    &   & $x=0$ & $x=0.1$ & $x=0.19$ \\
  \hline
  $a~(\AA)$      &   & $4.8300(5)$& $4.8120(4)$ & $4.7946(6)$ \\
  $b~(\AA)$      &   & $5.7597(6)$& $5.7645(5)$ & $5.7694(7)$ \\
  $c~(\AA)$      &   & $4.9977(5)$ & $4.9838(4)$ & $4.9708(7)$ \\
  $V~(\AA^3)$    &   & $139.009(3)$ & $138.208(2)$ & $137.453(3)$ \\
 $\beta(^\circ)$ &   & $91.140(7)$ & $91.370(7)$ & $91.579(9)$ \\
 \hline
  Atoms &   &   &   &   \\
  Mn/Cu & $y/b$  & $0.6861(4)$  & $0.6854(6)$  & $0.6875(1)$  \\
        & $B_{iso}~(\AA^{2})$ & $0.5120(47)$ & $0.4970(57)$  & $0.429(11)$ \\\cline{2-5}
    W   & $y/b$  & $0.1815(3)$  & $0.1806(36)$  & $ 0.1806(3)$  \\
        & $B_{iso}~(\AA^{2})$ & $0.5070(38)$  & $0.5550(38)$ & $0.6620(47)$ \\
  \hline
  \multicolumn{2}{l} {Global weighted $\chi^{2}~(\%)$} & $5.73$ & $8.2$ & $8.4$ \\
  \hline \hline
\end{tabular*}
\end{table}

Magnetization measurements of all samples were performed under magnetic fields of 1~kOe. Thermal evolution of magnetic susceptibility of the samples at low temperature is presented in figure~\ref{Merged_magnetization_properties}(a). From the magnetic susceptibility data of samples x=0.05, 0.1 and 0.15 only one magnetic ordering temperature is discernible around 14~K and for sample x=0.19 two anomalies are discernible. The thermal evolution of inverse susceptibility obtained from the field-cooled magnetization were fitted with Curie-Weiss law as shown in figure~\ref{Merged_magnetization_properties}(b). Inverse susceptibility of Mn$_{1-x}$Cu$_{x}$WO$_{4}$ follows Curie-Weiss law down to $\sim75$~K, below which it deviates from the fitted curve and shows a marked deviation below $\sim15$~K which corresponds to $T_{\mathrm {N1}}$. The deviation of inverse susceptibility well above ordering temperatures indicates the presence of short-range spin fluctuations above $T_{\mathrm {N}}$. Thermal evolution of Curie-Weiss temperature ($\Theta_{\mathrm{CW}}$) and the frustration parameter calculated as $f=|\Theta_{\mathrm{CW}}|/T_{\mathrm N}$ is presented in figure~\ref{Merged_magnetization_properties}(c) as a function of composition. Indeed MnWO$_4$ has been known to be a moderately spin frustrated system with the frustration parameter, $f=|\Theta_{\mathrm{CW}}|/T_{\mathrm N}\approx4.9$, where $\Theta_{\mathrm{CW}}$ is approximately $-71$~K and the N\'{e}el temperature $T_{\mathrm N}=13.5$~K~\cite{HDachs_7SSC1969Zur, AHArkenbout_74PRB2006ferroelectricity}. From figure~\ref{Merged_magnetization_properties}(c) it is clear that Cu doping increases the Curie-Weiss temperature and reduces frustration.

\begin{figure}[!t]
\centering
\includegraphics[scale=0.4]{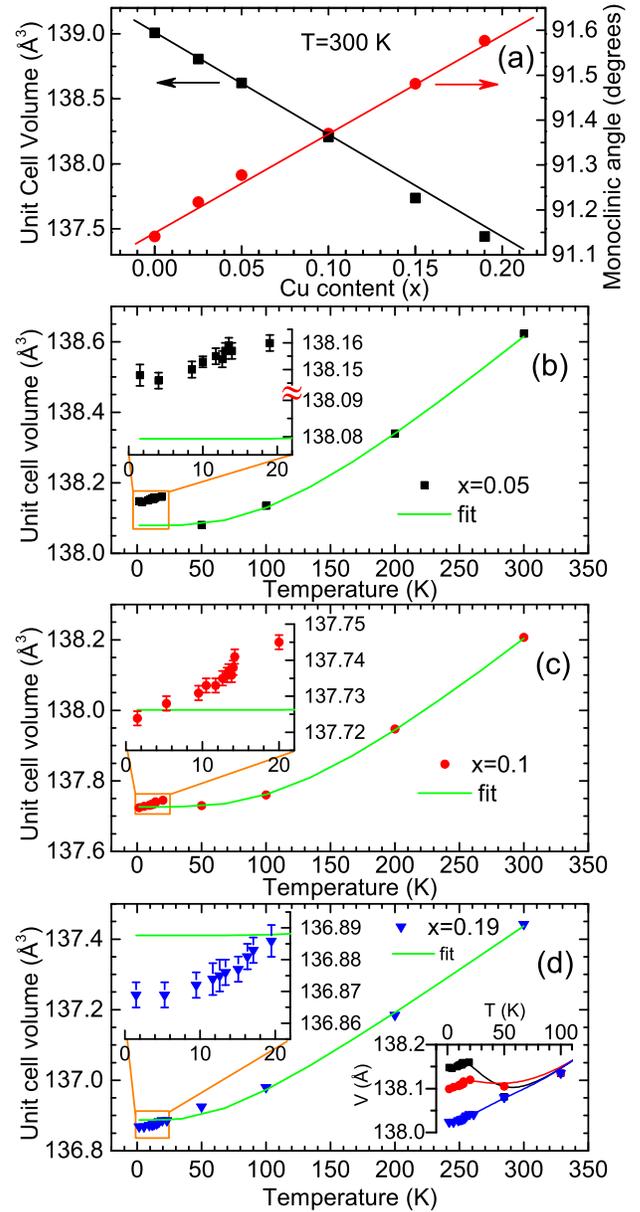}
\caption{(Colour online) (a) Unit cell volume and monoclinic angle as a function of Cu concentration at temperature 300~K, lines are only guides. (b)-(d) The temperature dependence of unit cell volume for the Cu compositions x=0.05, 0.1 and 0.19. The curves are fits to equation~\ref{VolumeDebyeFit} with the Gr\"{u}neisen approximation for the zero pressure equation. Insets in (b)--(d) show the deviation of unit cell volume from the fitted data at low temperatures. In the second inset of (d) the unit cell volumes are plotted together, to compare the magnitude of negative thermal expansion close to ordering temperature the y axes of x=0.1 and 0.19 are scaled by adding 0.375~{$\AA^3$} and 1.155~{$\AA^3$} respectively to match the value of the sample x=0.05 at 100~K. The lines are guides to eyes.}
\label{MonoclinicAngle_UnitCellVolume_DebyeFit}
\end{figure}

\begin{figure*}[!htb]
\centering
\includegraphics[scale=0.34]{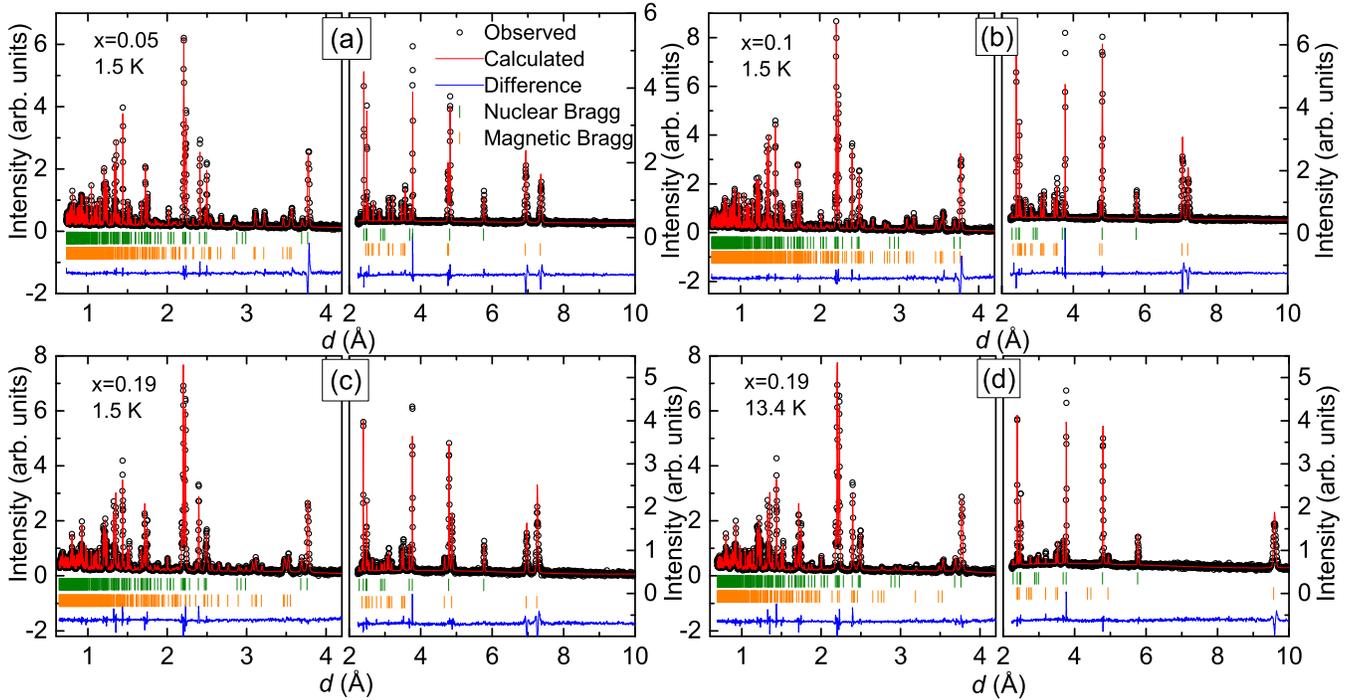}
\caption{(Colour online) Observed and calculated diffractions patterns and their difference for (a) Mn$_{0.95}$Cu$_{0.05}$WO$_4$ at 1.5~K (b) Mn$_{0.9}$Cu$_{0.1}$WO$_4$ at 1.5~K (c) Mn$_{0.81}$Cu$_{0.19}$WO$_4$ at 1.5~K and (d) Mn$_{0.81}$Cu$_{0.19}$WO$_4$ at 13.4~K. Circles are the measured intensities and the curve is the calculated pattern. Top and bottom vertical bars mark the positions of the nuclear and magnetic Bragg reflections respectively. Bottom curve is the difference between the measured and calculated patterns.}
\label{MnCuW_MagneticRefinedPatterns}
\end{figure*}

The temperature evolution of diffraction patterns in a selected $d$-space is presented in figure~\ref{2D_Intensity_Maps}. A magnetic phase transition is discernible based on the new incommensurate Bragg peaks below 13.5~K and 14~K in Mn$_{0.95}$Cu$_{0.05}$WO$_{4}$ and Mn$_{0.9}$Cu$_{0.1}$WO$_{4}$, respectively. In the case of Mn$_{0.81}$Cu$_{0.19}$WO$_{4}$ two transitions were observed around 17~K and 11.5~K. The fundamental crystal structure of all compositions is monoclinic with space group $P2/c$. The unit cell volume and monoclinic angle at 300~K is presented in figure~\ref{MonoclinicAngle_UnitCellVolume_DebyeFit}(a), the refined structure parameters including errors are tabulated in table~(\ref{Refinement_results_three_compositions}). With the increase in Cu concentration, a decrease in the unit cell volume and increase in the monoclinic angle was observed. The volume change is in accordance with the Vegard's law, lattice volume decreases as x increases, because the ionic radius of Cu$^{2+}$ is smaller than Mn$^{2+}$~\cite{RDShannon_32ActCry1976revised}. The temperature dependence of unit cell volume is shown in figure~\ref{MonoclinicAngle_UnitCellVolume_DebyeFit}(b)-(d). All three compositions presented here show a negative thermal expansion, an increase in volume with the decrease in temperature below 50~K. This effect seems to halt around 20~K and the volume starts to shrink below magnetic ordering temperature. The effect is more pronounced in lower Cu concentration and diminishes significantly with Cu doping as shown in the lower inset of figure~\ref{MonoclinicAngle_UnitCellVolume_DebyeFit}d. The anomalous behavior of the temperature variation of unit cell volume is due to the magneto-elastic effect associated with the antiferromagnetic transition at the N\'{e}el temperature. In order to study the spontaneous magnetostriction it is necessary to determine the temperature variation of the lattice parameters and the unit cell volume in the absence of magnetism. One way to determine the background temperature variation of the lattice parameter and unit cell volume is to extrapolate the paramagnetic high temperature region to low temperature by fitting with a polynomial function~\cite{TChatterji_21JPCM2009neutron}. This is only an approximation which works in some cases but involves some uncertainty. Alternatively, we used the Gr\"{u}neisen approximation for the zero pressure equation of state, in which the effects of thermal expansion are considered to be equivalent to elastic strain~\cite{DCWallace_book1998thermodynamics}. Within this approach the temperature dependence of the volume can be described by,
\begin{equation}
\label{VolumeDebyeFit}
V(T) = \gamma U(T)/{B_0} + {V_0}
\end{equation}
\noindent
where, $\gamma$ is a Gr\"{u}neisen parameter, $B_0$ is the bulk modulus and $V_0$ is the volume at $T=0$~K in the absence of magnetoelastic effect. By adopting the Gr\"{u}neisen approximation, the internal energy $U(T)$ is given by
\begin{equation}
\label{Grueneisen spproximation_free energy}
U(T) = 9N{k_{\rm{B}}}T{\left( {\frac{T}{{{\theta _{\rm{D}}}}}} \right)^3}\int_0^{{{{\theta _{\rm{D}}}} \mathord{\left/
 {\vphantom {{{\theta _{\rm{D}}}} T}} \right.
 \kern-\nulldelimiterspace} T}} {\frac{{{x^3}}}{{{{\mathop{\rm e}\nolimits} ^x} - 1}}} {\rm{d}}x
\end{equation}
\noindent
where $N$ is the number of atoms in the unit cell, $k_{\mathrm{B}}$ is Boltzmann's constant and $\theta _{\mathrm{D}}$ is the Debye temperature. By fitting the unit cell volume in the paramagnetic state we can get the physical parameters $\theta _{\mathrm{D}}$ and $V_0$. From the present fitting procedure it was not possible to determine $\gamma$ and $B_0$ but the ratio of $\gamma/ B_0$ was set as variable. The result of the fit is shown as a green solid line in figure~\ref{MonoclinicAngle_UnitCellVolume_DebyeFit}(b-d). Remarkably the fitted curves deviate from the experimental data at around 50~K much above the magnetic ordering temperature, where the unit cell volume undergoes a negative thermal expansion. Below magnetic ordering temperature $T_{\mathrm{N1}}$ the negative volume effect seized and the unit cell volume is decreased with temperature. This is inferred as a clear indication of presence of magnetoelastic or magnetovolume effects in these system as well as underlying frustration, though the negative thermal expansion is significantly small for x=0.19 compound. Temperature evolution of lattice parameters are similar to Co doped compound in which complex magnetic phase diagram is attributed to lattice changes~\cite{YSSong_22PRBStabilization}.

\begin{table}[t]
	\caption{The little group of ${\bf k}$=$(\alpha,1/2,\gamma)$=$(-0.2183,~1/2,~0.476)$ in P2/c is ${\bf{G_k}}$=${\bf P}_c$ (with co-set representatives with respect to the translation group: 1,c). The corresponding IRs are one-dimensional $\Gamma_1(1,c)$=$(1,\mathit{\upvarepsilon})$ and $\Gamma_2(1,c)$=$(1,-\upvarepsilon)$, with --$\upvarepsilon$=exp(${\uppi}$i$\upgamma$). The basis vectors of the IRs of  ${\bf{G_k}}$ are given below for the atoms Mn/Cu in the primitive unit cell numbered as: 1(1/2,~$y$,~1/4) and 2(1/2,~1$-y$,~3/4) related by $c$-glide plane:  $c$~($x$,~$-y$+1,~$z$+1/2), respectively.}
	\label{basis_vector_table}
	\begin{tabular}{ccc|cccccc}
		\hline\hline
		IR  &  BV  &  Atom & \multicolumn{6}{c}{BV components}\\
		&      &             &$m_{\|a}$ & $m_{\|b}$ & $m_{\|c}$ &$im_{\|a}$ & $im_{\|b}$ & $im_{\|c}$ \\
		\hline
		$\Gamma_{1}$ & $\bf \psi_{1}$ &      1 &      1 &      0 &      0 &      0 &      0 &      0  \\
		&              &      2 &   -0.075 &      0 &      0 &   0.997 &      0 &      0  \\
		& $\bf \psi_{2}$ &      1 &      0 &      1 &      0 &      0 &      0 &      0  \\
		&              &      2 &      0 &  0.075 &      0 &      0 &  -0.997 &      0  \\
		& $\bf \psi_{3}$ &      1 &      0 &      0 &      1 &      0 &      0 &      0  \\
		&              &      2 &      0 &      0 &   -0.075 &      0 &      0 &   0.997  \\
		$\Gamma_{2}$ & $\bf \psi_{4}$ &      1 &      1 &      0 &      0 &      0 &      0 &      0  \\
		&              &      2 &  0.075 &      0 &      0 &  -0.997 &      0 &      0  \\
		& $\bf \psi_{5}$ &      1 &      0 &      1 &      0 &      0 &      0 &      0  \\
		&              &      2 &      0 &   -0.075 &      0 &      0 &   0.997 &      0  \\
		& $\bf \psi_{6}$ &      1 &      0 &      0 &      1 &      0 &      0 &      0  \\
		&              &      2 &      0 &      0 &  0.075 &      0 &      0 &  -0.997  \\
		\hline \hline
	\end{tabular}
\end{table}

Representational analysis allows the determination of the symmetry-allowed magnetic structures that can result from a second-order magnetic phase transition, given the crystal structure before the transition and the magnetic propagation vector (${\bf k}$) of the magnetic ordering. These calculations were carried out using the program $BasIreps$ included in the $FullProf$ suite. First, the program $k-search$, also included in the $FullProf$ suite, is used to determine the magnetic propagation vector at different temperatures. For x=0.05 and 0.1 the magnetic propagation vector was found to be ${\bf k}=(k_x,\frac{1}{2},~k_z)$ in whole temperature range. For x=0.19 the magnetic propagation vector in the temperature range 11.5-17~K was found to be $\mathbf {k}=({\textstyle{1 \over 2}}, 0, 0)$ and below 11.5~K it is ${\bf k}=(k_x,\frac{1}{2},~k_z)$. While the magnetic propagation vector determines the modulation going from one unit cell to another, magnetic symmetry analysis is needed to determine the coupling between the symmetry related magnetic sites within one crystallographic unit cell. $BasIreps$ is used to compute all the allowed symmetry couplings in the form of irreducible representations and their respective basis vectors. The Mn/Cu at the site 2f in the crystallographic space group P2/c, for the incommensurate magnetic propagation vector ${\bf k}=(k_x,\frac{1}{2},~k_z)$, is found to have two possible irreducible magnetic representations (IR) each having three basis vectors (BV). The computed results for the x=0.05 compound at 1.5~K with the propagation vector ${\bf k}=(-0.218,\frac{1}{2},0.476)$ are presented in table~(\ref{basis_vector_table}). All possible combinations of the two allowed irreducible representations were tested against the measured data. Rietveld refinements clearly showed that only with the IR $\Gamma_{1}$ (with real and imaginary components) a successfull refinement of the data can be obtained. The propagation vector and the refined coefficient of basis vectors, C1,~C2 and C3, for x=0.05, 0.1 and 0.19 at 1.5~K is presented in table~(\ref{Refined coefficients of basis vectors}). It should be noted that C1 and C3 are real coefficients while C2 is a pure imaginary coefficient. For x=0.19 in the temperature range 11.5-17~K, with  $\mathbf {k}=({\textstyle{1 \over 2}}, 0, 0)$, four one-dimensional IRs were computed, for $\Gamma_1$ and $\Gamma_2$ two BVs are allowed and for $\Gamma_3$ and $\Gamma_4$ only one BV is allowed. The Shubnikov groups (SG) of each IRs have the same symbol P$_{a}2/c$ (in Belov-Neronova-Smirnova notation), but they correspond to different magnetic Wyckoff positions and origin shifts~\cite{JMPerezMato_45AnuRevMaterRes2015SymmetryBased}. The magnetic moments of the two atoms (1)($x$,~$y$,~$z$) and (2)($-x$+1,~$-y$+1,~$-z$+1) in the paramagnetic unit cell have the following configurations for eanch representation: $\Gamma_1$:1$(m_x,0,m_z)$, 2$(-m_x,0,-m_z)$; $\Gamma_2$:1$(m_x,0,m_z)$, 2$(m_x,0,m_z)$; $\Gamma_3$:1$(0,m_y,0)$, 2$(0,-m_y,0)$ and $\Gamma_4$:1$(0,m_y,0)$, 2$(0,-m_y,0)$. Only the representation described by $\Gamma_2$ (SG-P$_{a}2/c$, Wyckoff site $4f$ in the doubled unit cell) provides a successfull refinement of the data with $m_x$=1.24(7)~$\mu_{\mathrm B}$ and $m_z$=1.28(8)~$\mu_{\mathrm B}$ for $T$=13.5~K.

\begin{table}[!t]
	\caption{Refined unit cell parameters, magnetic propagation vector and coefficients of basis vectors for Mn$_{1-x}$Cu$_x$WO$_4$ at 1.5~K.}
	\label{Refined coefficients of basis vectors}
	\begin{tabular*}{\columnwidth}{@{\extracolsep{\fill}}ccccc}
		\hline \hline
		&   &  \multicolumn{3}{c}{Cu content (x)}\\ \cline{3-5} %% This will add a horizontal line along the cells 3 to 5
		&   & $x=0.05$ & $x=0.1$ & $x=0.19$ \\
		\hline
		\multicolumn{3}{l}{Unit cell parameters}   &   &   \\
		      & $a~(\AA)$  & $ 4.8145(6)$& $4.8045(4)$ & 4.7856(8) \\
		      & $b~(\AA)$  & $5.7565(9)$ & $5.7583(6)$ & 5.7619(8) \\
		      & $c~(\AA)$  & $4.9860(8)$ & $4.9788(5)$ & 4.9653(9) \\
		      & $\beta(^\circ)$  & $91.20(1)$ & 91.30(9) & 91.49(6) \\
		\hline
		Mn/Cu & $y/b$  & $0.6841(6)$  & 0.6840(6)  &  0.6857(2) \\
		      & $B_{iso}~(\AA^{2})$ & $0.274(6)$   & 0.184(2)  & 0.203(1)  \\
		\hline
		\multicolumn{4}{l}{Components of propagation vector ($k_y=\textstyle{1 \over 2}$)}     &   \\
				           & $k_x$ & -0.218(3)  & -0.221(6)  & -0.223(4)   \\
				           & $k_z$ & 0.476(1)   & 0.494(4)  & 0.526(4)  \\
        \hline
		 \multicolumn{3}{l}{Coefficients of basis vectors}  &   &   \\
			   & C1(real)  & $3.41(1)$ & 3.28(1)  & 2.79(9) \\
			   & C2(imaginary)  & $-3.96(2)$& -4.08(9) & -3.62(4) \\
			   & C3(real)  & $2.95(1)$ & 2.91(1)  & 2.32(3) \\
		\hline \hline
	\end{tabular*}
\end{table}

\begin{figure}[!t]
\centering
\includegraphics[scale=0.42]{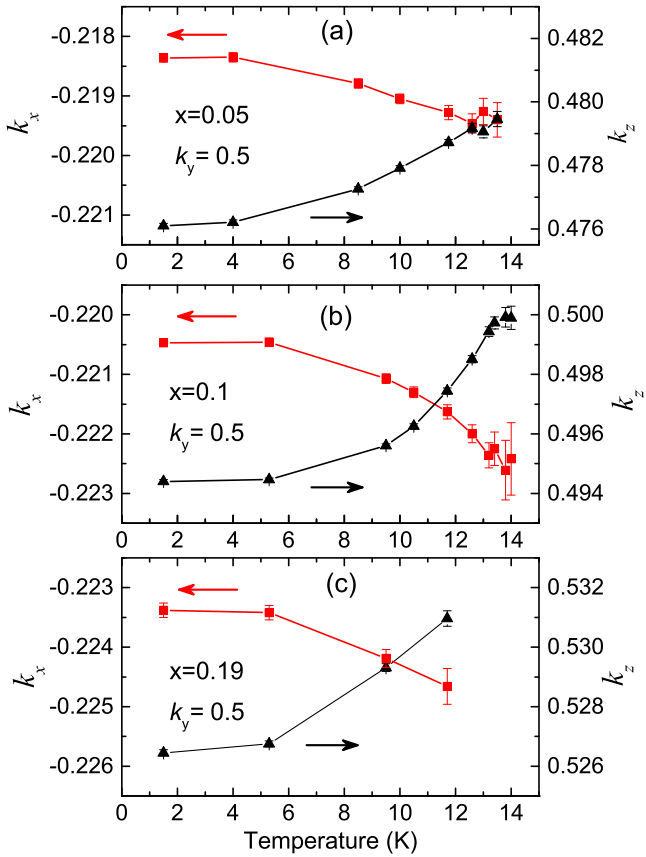}
\caption{(Colour online) Temperature variation of components of incommensurate propagation vector along $x$~($k_x$) and $z$~($k_z$) axes for (a) Mn$_{0.95}$Cu$_{0.05}$WO$_4$, (b) Mn$_{0.9}$Cu$_{0.1}$WO$_4$ and (c) Mn$_{0.81}$Cu$_{0.19}$WO$_4$. Component of propagation vector along $y$ is $k_y = 0.5$.}
\label{Evolution_of_Propagation_vectors}
\end{figure}

\begin{figure*}[!htb]
	\centering
	\includegraphics[scale=0.39]{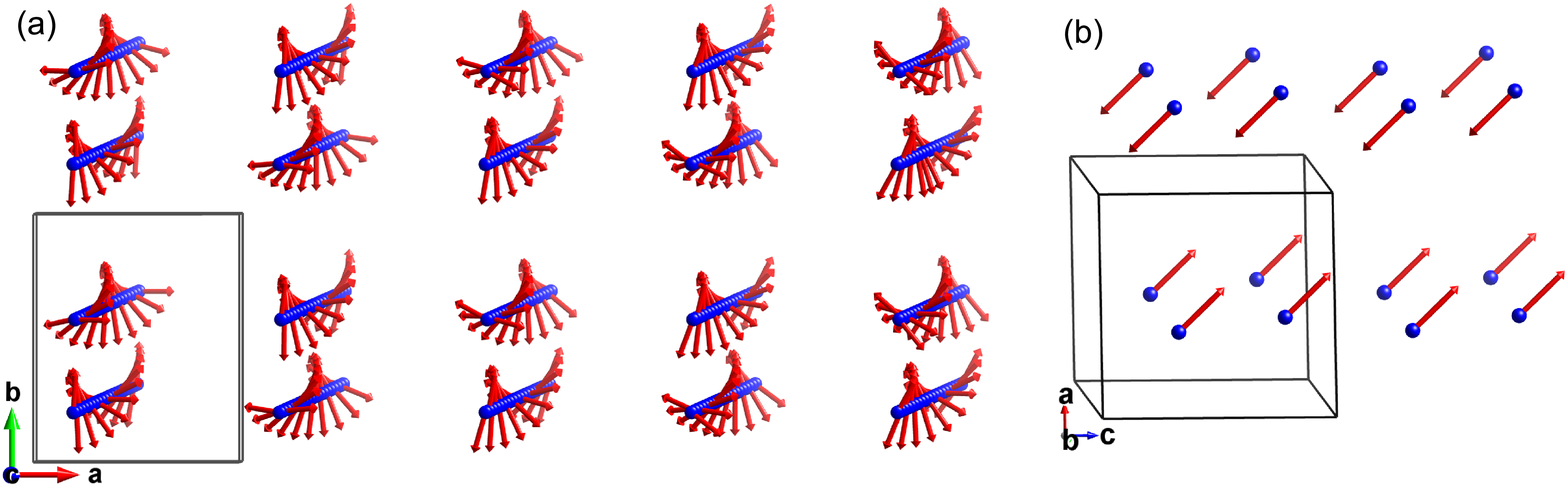}
	\caption{(Colour online) (a) ICM structure of Mn$_{0.81}$Cu$_{0.19}$WO$_4$ with propagation vector $\mathbf{k}=(-0.225, \textstyle{1 \over 2}, 0.531)$, the number of visible unit cells along a, b and c direction are, 5, 2 and 20, respectively.  (b) High temperature CM structure of Mn$_{0.81}$Cu$_{0.19}$WO$_4$ with propagation vector (0.5, 0, 0), two unit cells along all three axes are shown. Gray box indicate one unit cell.}
	\label{Magnetic_Structures}
\end{figure*}

Refined NPD patterns are presented in figure~\ref{MnCuW_MagneticRefinedPatterns}(a-d). From the magnetic structure refinements {Mn$_{0.95}$Cu$_{0.05}$WO$_4$} was found to order at $\sim13.5$~K, with the incommensurate propagation vector $\mathbf {k}=(-0.218,\frac{1}{2},0.477)$. The temperature dependence of components of incommensurate propagation vector ${\bf k}=(k_x,\frac{1}{2},~k_z)$ is presented in figure~\ref{Evolution_of_Propagation_vectors}(a). With decrease in temperature a distinct change in $k_x$ and $k_y$ is observed. In case of MnWO$_4$ the magnetic phases AF2 and AF3 are incommensurate with similar wave vector ${\bf k}=(-0.214,~\frac{1}{2},~0.457)$. Only AF2 phase with elliptical spin arrangement is ferroelectric which can be explained by spatial inversion symmetry breaking spiral spin structure~\cite{KTaniguchi_97PRL2006ferroelectric}. It should be noted that in the case of {Mn$_{0.95}$Cu$_{0.05}$WO$_4$} in the whole temperature range the structure is found to be similar to that of AF2 in the parent compound. The magnetic structure in case of Mn$_{0.9}$Cu$_{0.1}$WO$_4$ is similar to that of {Mn$_{0.95}$Cu$_{0.05}$WO$_4$} but with a modified propagation vector $\mathbf {k}=({-0.224, \textstyle{1 \over 2}}, 0.5)$ close to 14~K. The evolution of propagation vector with temperature in {Mn$_{0.9}$Cu$_{0.1}$WO$_4$} is presented in figure~\ref{Evolution_of_Propagation_vectors}(b). Striking change in propagation vector close to ordering temperature $T_\mathrm{N1}$ in this case indicates that with increased Cu content of x=0.1 the propagation vector along z-direction is nearly commensurate. From our powder diffraction measurements for x=0.05 and 0.1 compounds we don't see any significant change associated with the transition from $T_{\mathrm{N1}}$ (AF3) to $T_{\mathrm{N2}}$ (AF2) as seen from specific heat measurements. Considering very narrow temperature range between these two transitions it might be difficult to clarify this with our bulk powder measurements. We expect that for x=0.05-0.15, the magnetic ordering in the temperature range $T_{\mathrm{N2}} {<} T {<} T_{\mathrm{N1}}$ should be collinear incommensurate phase as in MnWO$_4$~\cite{KTaniguchi_97PRL2006ferroelectric} with magnetic propagation vector similar to AF2 phase. Further studies on single crystals with polarized neutron diffraction with smaller temperature steps will be a best tool for the detailed investigation of this structure. 

With further increase in doping in case of Mn$_{0.81}$Fe$_{0.19}$WO$_4$ we observe a commensurate magnetic (CM) structure at $T_\mathrm{N1}$=17~K with $\mathbf {k}=({\textstyle{1 \over 2}}, 0, 0)$ which is similar to AF4 phase in parent compound. Below 11.5~K it undergoes another magnetic phase transition to AF2 phase with propagation vector $\mathbf{k}=(-0.225, \textstyle{1 \over 2}, 0.531)$ which is modulated with temperature as shown in figure~\ref{Evolution_of_Propagation_vectors}c. The incommensurate cycloidal (AF2) and the commensurate collinear (AF4) magnetic structure of Mn$_{0.81}$Cu$_{0.19}$WO$_4$ is presented in figure~\ref{Magnetic_Structures}(a) and (b), respectively. The incommensurate structures for lower doping systems is quite similar to the one presented in figure~\ref{Magnetic_Structures}(a).

\begin{figure}[!b]
	\centering
	\includegraphics[scale=0.45]{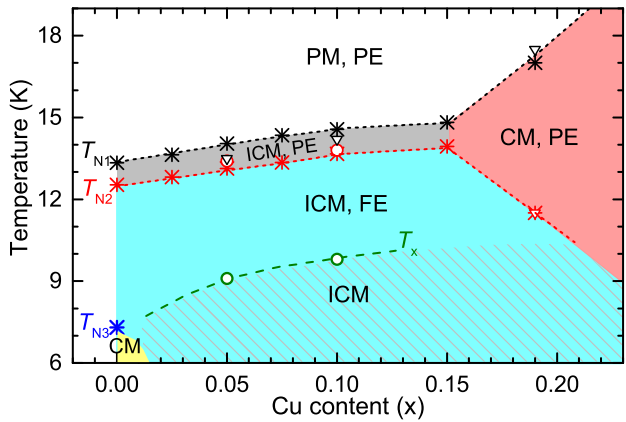}
	\caption{Tentative phase diagram of Mn$_{1-x}$Cu$_{x}$WO$_4$ with phase boundaries obtained from specific heat (asterisk), NPD (inverted triangle) and dielectric (circle) measurements. The symbols in black, red and green corresponds to the transitions $T_{\mathrm {N1}}$, $T_{\mathrm{N2}}$ and $T_{\mathrm {X}}$, respectively. Blue asterisk indicates the transition from AF2 to AF1 phase in MnWO$_4$. (PM--paramagnet, CM--commensurate magnet, ICM--incommensurate magnet, FE--ferroelectric and PE--paraelectric)}
	\label{PhaseDiagram}
\end{figure}

From our comprehensive study of Mn$_{1-x}$Cu$_{x}$WO$_4$ we are able to construct a tentative magnetic phase diagram as shown in figure~\ref{PhaseDiagram}. The magnetic phase diagram of Cu doped compound found to be much simpler than that of Co doped compound~\cite{YSSong_22PRBStabilization, FYe_86PRB2012Magnetic} but very similar to Zn doped compound~\cite{RPChaudhury_83PRB2011robust}. With higher doping concentration a collinear magnetic structure is stabilized at higher temperatures. This is attributed to weakening of spin-phonon coupling and hence lower frustration leading to a simpler magnetic ordering. From the neutron diffraction measurement it is clear that the low temperature phase below $T_\mathrm{X}$ which is observed from dielectric measurements is incommensurate cycloidal phase. Magnetic structure refinements confirmed that the magnetic phase below $T_\mathrm{X}$ (region marked with gray lines in figure~\ref{PhaseDiagram}) is not associated with the transition from cycloidal structure with magnetic vectors ${\bf k}=(k_x,0.5,k_z)$ to collinear structure with magnetic vector ${\bf k}=(0.25,0.5,-0.5)$ as seen in MnWO$_4$~\cite{AHArkenbout_74PRB2006ferroelectricity}. This leads to the inference that below $T_\mathrm{X}$ the magnetic structure undergoes a temperature induced spin flip transition with similar magnetic propagation vectors which is indistinguishable from powder diffraction measurements. The suppression of low temperature collinear phase can be attributed to extremely sensitive exchange coupling and anisotropy constants with respect to perturbations~\cite{MKenzelmann_74PRB2006Field,RPChaudhury_83PRB2011robust}. In the present case chemical doping plays the role of perturbations. In a recent report based on magnetization, specific heat and ferroelectric polarization measurements, Liang et al showed that by the substitution of lower spin (1/2) Cu$^{2+}$ for the higher spin (5/2) Mn$^{2+}$ ion the multiferroic phases of MnWO$_4$ are strongly affected~\cite{KCLiang_470PRB2014control}. The Cu substitution will introduce a low spin with different exchange coupling and anisotropy constants affecting the magnetic and ferroelectric states. This leads to the stabilization of ferroeletric spin spiral phase at low temperatures with increasing Cu content. The microscopic exchange interactions can be obtained through inelastic neutron scattering (INS) experiments investigating the magnetic excitations. According to INS experiments on MnWO$_4$, to explain the magnetic excitation spectrum, up to 11 different exchange pathways were required to fit the data proving the three dimensional character of magnetic fluctuations~\cite{HEhrenberg_11JPCM1999}. This three dimensional nature explains the robustness of cycloidal spiral order in Cu-doped MnWO$_4$, since percolation threshold for site dilution is much lower than for two-dimensional systems~\cite{PKharel_89PhilsoMag2009}. Based on a semiphenomenological Landau theory, authors in~\cite{SMatityahu_85PRB2012Landau} clarified the effect of different dopants on the phase diagram of Mn$_{1-x}$M$_{x}$WO$_4$ ($M$=Fe,Zn,Mg). The origin of complex phase diagrams in these compounds is attributed to competition between different superexchange interactions with contrasting behavior of doping with different ions. We expect that the theoretical analysis presented in~\cite{SMatityahu_85PRB2012Landau} should be compatible for Mn$_{1-x}$Cu$_{x}$WO$_4$ as well. The temperature induced spin-reorientation remains to be unique to the present compound which requires further scrutiny.

\section{IV. Conclusions}

From our detailed investigation of the Mn$_{1-x}$Cu$_{x}$WO$_4$ for ${0}\leq{x}\leq{0.19}$ we have shown that substitution of Cu for Mn results in a reduction of the frustration. Also a reduction in negative thermal expansion with the increased Cu doping was observed which hints to a reduction in spin-phonon coupling with the higher Cu content. Temperature and doping dependence of lattice parameters establish a strong dependence of magnetic structure on lattice changes. Both $T_{\mathrm {N1}}$ and $T_{\mathrm{N2}}$ increased with higher Cu content. This is in contrast to Mn$_{1-x}$Zn$_{x}$WO$_4$~\cite{RPChaudhury_83PRB2011robust}. The presence of third transition $T_{\mathrm{X}}$ is unique to the present compound. We note again, our NPD data confirms that the origin of $T_{\mathrm{X}}$ is not ICM to CM observed in MnWO$_4$ at $T_{\mathrm{N3}}$. A possible origin of this phase transition is the temperature induced spin reorientation. Further single crystal neutron diffraction and electric polarization measurements are desirable to shed light on the nature of ferroelectric and magnetic ordering below $T_{\mathrm{X}}$ and in the region between $T_{\mathrm{N1}}$ and $T_{\mathrm{N2}}$.

\section{ACKNOWLEDGMENTS}
We thank the expert assistance of A. Huq, SNS, Oak Ridge National Laboratory, during the NPD measurements. The authors gratefully acknowledge the financial support provided by JCNS to perform the neutron scattering measurements at the Spallation Neutron Source (SNS), Oak Ridge, USA. Part of the research conducted at SNS was sponsored by the Scientific User Facilities Division, Office of Basic Energy Sciences, US Department of Energy. The work at the University of Augsburg was supported by the Deutsche Forschungsgemeinschaft via the Transregional Collaborative Research Center TRR 80.

\end{document}